
\documentstyle{amsppt}


\pagewidth{6.4 truein}
\pageheight{8.6 truein}
\hfuzz=20pt


\TagsOnRight

\topmatter
\title Evaluating the Crane-Yetter Invariant \endtitle
\author Louis Crane \\ Louis H. Kauffman \\ David Yetter \endauthor
\address Louis Crane, Department of Mathematics, Kansas State
         University, Manhattan, KS  66506-2602, USA \endaddress
\address Louis H. Kauffman, Department of Mathematics, University of
         Illinois, Chicago, IL  60680, USA \endaddress
\address David Yetter, Department of Mathematics, Kansas State
         University, Manhattan, KS  66506-2602, USA \endaddress


\endtopmatter

\vglue .2in

\document

\head I. Introduction \endhead

The purpose of this paper is to give an explicit formula for the
invariant of  $4$-manifolds introduced by Crane and Yetter in [CR93].
For a closed  $4$-manifold  $W$, this invariant will be denoted herein
by  $CY(W)$.  Our main result is the following theorem, proved in
section 4.

\vskip .2in

\proclaim{Theorem} Let  $W$  be a closed  $4$-manifold.  Let  $\sigma
(W)$  denote the signature of  $W$, $\chi (W)$  denote the Euler
characteristic of  $W$  and  $CY(W)$  denote the Crane-Yetter invariant
of  $W$.  Let values  $N$  and  $\kappa$  be defined as in the beginning
of section 4.  Then  $CY(W) = \kappa^{\sigma (W)} N^{\chi (W)/2}$.
\endproclaim

\vskip.2in

This result is of general interest because {\it it expresses the
signature of a  $4$-manifold in terms of local combinatorial data}
(these data produce the state summation  $CY(W)$  in terms of a
triangulation of  $W$).  Our result should be compared with the work of
Gelfand and Macpherson [GM92] where the Pontrjagin classes (and hence
the signature) are produced by a combination of subtle combinatorics and
geometric topology.  Here we give a formula for the signature in terms
of a topological quantum field theory that is based on  $SU(2)q$  and on
$q$-deformed spin networks.

It should also be mentioned that the invariant  $CY(W)$  is a rigouous
version of ideas of Ooguri [OG92].  It is of interest to examine the
implications of our result for the physics that is inherent in Ooguri's
work.  In section 2 we recall the definition of the Crane-Yetter
invariant.  In section 3, $CY(W)$  is reformulated in terms of
Temperley-Lieb recoupling theory.  The Theorem is proved in section 4.

Louis Crane wishes to acknowledge support by NSF Grant DMS-9106476.
Louis Kauffman wishes to acknowledge support by NSF Grant DMS-9205277
and the Program for Mathematics and Molecular Biology of the University
of California at Berkeley, Berkeley, CA.

\vskip .2in

\head II. A Concise Description of the Invariant  $CY(W)$ \endhead

Let  $W = W^4$, a closed  $4$-manifold.  Let  $CY(W)$  denote the
Crane-Yetter invariant of  $W$, as defined in [CY93].  The formula for
this invariant, in terms of a triangulation of  $W$, is a state
summation over colorings from the index set  $\{ 0,1,2,\ldots,r - 2\}$
of the two dimensional faces and three dimensional simplices of the
triangulation of  $W$.  Here we use integer labellings for convenience.
In [CY93] the labellings are by half integers, but the formulas are
equivalent.  Thus the invariant is a function of the integers  $r =
3,4,\ldots$.

To each colored face of the triangulation is assigned the quantum
integer (quantum dimension)
$$
\Delta (\text{face}) = \Delta (i) = (-1)^i [q^{i+1} - q^{i-1}]/[q -
q^{-1}] \quad\text{where}\quad q = \exp (i \pi/r)
$$
where  $i$  denotes the color assigned to that face.

To each colored tetrahedron is assigned  $\Delta (\text{tet}) = \Delta
(i)$, where  $i$  is the color assigned to that tetrahedron.

To each  $4$-simplex is assigned the ``15-j symbol''  $\Phi$  (4 plex)
that is associated to the coloration of its boundary.  This 15-j symbol
is an evaluation of a network associated with the boundary of the
$4$-simplex, obtained by having two interconnected  $3$-vertices in each
tetrahedron, forming a network with four free ends corresponding to the
boundary of the tetrahedron.  These nets are then interconnected in the
pattern of the joinings of the faces of the tetrahedra in the boundary
of the  $4$-simplex.  In [CY93] a specific convention for forming this
net is given, and we refer to that paper for the details.  The
$3$-vertices and network evaluations are done by Crane and Yetter in the
Kirillov-Reshetikhin [KR88] diagrammatics for the recoupling theory of
$SU(2)q$.

The formula for the invariant  $CY(W)$  is then given as shown below.
$$
CY(W) = N^{n_0-n_1} \sum \prod \Delta \ (\text{face})\ \prod \Delta \
(\text{tet})^{-1} \prod \Phi (4\text{plex})
$$
The summation is over all colorings of the faces and tetrahedra from the
index set  $\{ 0,1,2,\ldots,r-2\}$.  The products are over all faces,
tetrahedra and  $4$-simplices respectively.  The values  $n_0$  and  $n_1$
are the number of  $0$-simplices and the number of  $1$-simplices in the
triangulation.  The value  $N$  is equal to the sum of the squares of
the quantum dimensions, and it has the specific value  $-2r/(q -
q^{-1})^2$:
$$
N = \sum_i \Delta (i)^2 = -2r/(q - q^{-1})^2.
$$
This completes our description of the invariant  $CY(W)$.

\vskip .2in

\head III. Translating  $CY(W)$  into the Temperley-Lieb Format \endhead

In order to prove our result about the evaluation of  $CY(W)$  it is
useful to translate the state sum into the language of the
Temperley-Lieb version of the recoupling theory.  This recoupling theory
is explained in detail in [KL93], and expositions of it are given in
[K91] and [K92].  We shall refer to Temperley-Lieb recoupling as the
{\it TL theory}, and to Kirillov-Reshetikhin recoupling as the {\it KR
theory}.

The TL theory is rooted in combinatorics of link diagrams, and it is a
direct generalization ($q$-deformation) of the Penrose spin network
theory.  Its advantage for us here is that there is no dependence in the
diagrammatics of the TL theory on maxima and minima or on the
orientation of the diagrams with respect to a direction in the plane.
Thus TL networks cana be freely embedded in handlebodies and
$3$-manifolds.

The basic information needed to transform KR nets into TL nets is the
relationship of their  $3$-vertices.  This is given by the formula below
where the subscripts KR and TL discriminate the vertices in question.
$$
[3-\text{vertex}/a,b;c]_{KR} = \left( \frac{\sqrt{\Delta
(c)}}{\sqrt{\theta (a,b,c)}}\right) [3-\text{vertex}/a,b,c]_{TL}
$$

Here  $\theta (a,b,c)$  is the TL evaluation of a theta net with edges
labelled  $a$, $b$, and  $c$  and  $[3 - \text{vertex}/a,b;c]_{(-)}$  is
the  $3$-vertex in the indicated recoupling theory with incoming edges
labelled  $a$  and  $b$  and outgoing edge labelled  $c$.

Note that the KR vertex is oriented with two legs up and one leg down.
The TL vertex does not have a dependence on the leg placement.  The
value of a closed loop labelled  $i = 0,1,\ldots,r - 2$  is  $\Delta
(i)$  in both theories.  We omit further details of the relationship
between TL and KR.

It is now a straightforward matter to translate the  $CY(W)$  into the
TL framework.  The result is as shown below where  $\phi$  denotes the
TL  $15-j$  symbol.  The TL  $15-j$  is described by exactly the same
diagram as the KR  $15-j$, but all its  $3$-vertices are of TL type.
$$
CY(W) = N^{n_0-n_1} \sum \prod \Delta \ (\text{face})\ \prod (\Delta \
(\text{tet}) \theta_1 \ (\text{tet})^{-1} \theta_2 \ (\text{tet})^{-1})
\prod
\phi
(4\text{plex})
$$

In this formula, $n_d$  is the number of  $d$-simplexes
in the triangulation of  $W$.  The sum is over all
colorings of the faces and tetrahedra of the triangulation.

$\theta_1 (\text{tet})$, $\theta_2 (\text{tet})$  are the two theta
evaluations assigned to each tetrahedron.  Each is of the form  $\theta
(a,b,c)$  where  $a$, $b$  and  $c$  are the colors assigned to one of
the  $3$-vertices in the net associated with this tetrahedron.  This
means that we can take  $a$  and  $b$  to be the colors of two (paired)
faces of the tetrahedron, and  $c$  to be the color associated with the
tetrahedron itself.

$\phi (4 \text{plex})$  is the TL evaluation of the  $15-j$  net
associated with each tetrahedron.

This completes our translation of the invariant  $CY(W)$  into the
Temperley-Lieb format.

\vskip .2in

\head IV. A Formula for  $CY(W)$ \endhead

In this section we use results of Justin Roberts [R93] to give an
explicit formula for  $CY(W)$  in terms of the Euler characteristic and
the signature of the closed  $4$-manifold  $W$.

(The reader should note that Roberts uses a notation  $CY(W)$, but that
his  $CY(W)$  is not precisely the Crane-Yetter invariant.  It differs
from it by a factor involving the Euler characteristic of the
$4$-manifold.)

We need the following values
$$
\aligned
\eta &= (q - q^{-1})/i \sqrt{(2r)} \\
\kappa &= \exp (i\pi (-3 - r^2)/2r) \exp (-i \pi/4).
\endaligned
$$
Note that  $N = \eta^{-2}$, where  $N$  is as defined in section 2.

Roberts [R93] proves that
$$
\kappa^{\sigma (W)} = \eta^{-n_0+n_1+n_2-n_3+n_4} S(W)
$$
where  $S(W) = \sum \prod \Delta \ (\text{face})\ \prod (\Delta \
(\text{tet})\ \theta_1 \ (\text{tet})^{-1} \theta_2 \ (\text{tet})^{-1})
\prod \phi (4\text{plex})$  and  $\sigma (W)$  denotes the signature of
the manifold  $W$.

Roberts' state summation  $S(W)$  has exactly the same form, except for
the power of  $N$, as our TL version of the Crane-Yetter invariant.  The
$15-j$  evaluations of Roberts involve orientation conventions that are
consistent with his use of TL networks in  $3$-dimensional handlebodies.
This means that it follows from Robert's work that different but
consistent conventions for the  $15-j$  symbols will lead to the same
results.  The Crane-Yetter convention in the TL format is one such
choice.
Therefore  $CY(W) = N^{n_0-n_1} S(W)$  by the results of section 3.

We can now prove the main theorem.

\vskip.2in

\proclaim{Theorem} Let  $W$  be a closed  $4$-manifold.  Let  $\sigma
(W)$  denote the signature of  $W$, $\chi (W)$  denote the Euler
characteristic of  $W$  and  $CY(W)$  denote the Crane-Yetter invariant
of  $W$.  Let values  $N$  and  $\kappa$  be defined as in the beginning
of section 4.  Then  $CY(W) = \kappa^{\sigma (W)} N^{\chi (W)/2}$.
\endproclaim

\vskip .2in

\demo{Proof}
$$
\aligned
\kappa^{\sigma (W)} &= \eta^{-n_0+n_1+n_2-n_3+n_4} N^{n_1-n_0} CY(W) \\
&= \eta^{-n_0+n_1+n_2-n_3+n_4} \eta^{-2n_1+2n_0} CY(W) \\
&= \eta^{n_0-n_1+n_2-n_3+n_4} CY(W) \\
&= N^{-\chi (W)/2} CY(W).  \qquad\text{Q.E.D.}
\endaligned
$$
\enddemo

\vskip .2in

\demo{Remark} Note that if we choose  $r$  greater than the number of
$2$-simplices in  $W$, then  $\sigma (W) < r$  and is therefore
determined by  $CY(W)$  and  $\chi (W)$  via the formula in the theorem.
Thus it is quite correct to say that the Theorem produces a
combinatorial formula for the signature of a compact  $4$-manifold in
terms of local data from the triangulation.
\enddemo


\vfill
\pagebreak

\vfill


\Refs
\widestnumber\key{PEN69}

\ref \key CR91 \by L. Crane \paper Conformal field theory, Spin
geometry and quantum gravity \jour Physics Letters B \vol 259 \issue 3
\pages 243-248 \yr 1991 \endref

\ref \key CR93 \by L. Crane and D. Yetter \paper A categorical
construction of  $4D$  topological quantum field theories \finalinfo (In
{\it Quantum Topology}, ed. by L.~Kauffman and R.~Baadhio), also
e-preprint hep-th 9301062.  \endref

\ref \key G92 \by I.M. Gelfand and R.D. Macpherson \paper A
combinatorial formula for the Pontrjagin classes \jour Bulletin of the
AMS \vol 26 \issue 2 \yr April 1992 \pages 304-308 \endref

\ref \key K90 \by L.H. Kauffman \paper Spin networks and knot
polynomials \jour Intl. J. Mod. Phys. A. \vol 5 \issue 1 \yr 1990 \pages
93-115 \endref

\ref \key K91 \bysame \book Knots and Physics \publ World
Scientific Pub. \yr 1991 \endref

\ref \key K92 \bysame \paper Map coloring, $q$-deformed spin networks,
and Turaev-Viro invariants for  $3$-manifolds \paperinfo In the
Proceedings of the Conference on Quantum Groups -- Como, Italy, June
1991 -- editor by M.~Rasetti, World Sci. Pub., \jour Intl. J. Mod.
Phys. B \vol 6 \issue 11 \& 12 \yr 1992 \pages 1765-1794 \endref

\ref \key KL93 \by L.H. Kauffman and S. Lins \paper Temperley Lieb
recoupling theory and invariants of  $3$-manifolds \toappear \endref

\ref \key KR88 \by A.N. Kirillov and N.Y. Reshetikhin \paper
Representations of the algebra  $U_q(sl_2)$, $q$-orthogonal polynomials
and invariants of links \paperinfo In {\bf Infinite Dimensional Lie
Algebras and Groups}.  \ed by V.G.~Kac \jour Adv. Ser. in Math. Phys.
\vol 7 \yr 1988 \pages 285-338 \endref

\ref \key OG92 \by H. Ooguri \paper Topological Lattice Models in Four
Dimensions \jour Mod. Phys. Lett. A \vol 7 \issue 30 \yr 1992 \pages
2799-2810 \endref

\ref \key PEN69 \by R. Penrose \book Angular momentum: An approach to
combinatorial space-time \bookinfo In {\bf Quantum Theory and Beyond}.
\ed T.A.~Bastin \publ Cambridge Univ. Press \yr 1969 \endref

\ref \key R93 \by J. Roberts \paper Skein theory and Turaev-Viro
invariants. \paperinfo (preprint 1993). \endref

\ref \key TV92 \by V.G. Turaev and O. Viro \paper State sum invariants
of  $3$-manifolds and quantum  $6j$  symbols \jour Topology \vol 31
\issue 4 \pages 865-902 \yr 1992 \endref

\endRefs
\enddocument